\documentclass[doublecol]{epl2}
\usepackage{amsmath}
\usepackage{amssymb}
\usepackage{epsfig}
\usepackage{color}
\usepackage{subfigure}
\usepackage{graphicx,amsmath}
\usepackage{graphicx}

\title{Identification of Strongly Correlated Spin Liquid  in Herbertsmithite}
\shorttitle{Identification of Spin Liquid}
\author{V. R. Shaginyan \inst{1,2}\thanks {Email:
\email{vrshag@thd.pnpi.spb.ru}} \and A. Z. Msezane \inst{2}\and K.
G. Popov\inst{3}\and G. S. Japaridze \inst{2} \and V. A.
Stephanovich\inst{4}} \shortauthor{V.R. Shaginyan \etal}
\institute{\inst{1} Petersburg Nuclear Physics Institute, Gatchina,
188300, Russia\\\inst{2} Clark Atlanta University, Atlanta, GA
30314, USA\\ \inst{3} Komi Science Center, Ural Division, RAS,
Syktyvkar, 167982, Russia\\ \inst{4} Opole University, Institute of
Physics, Opole, 45-052, Poland}


\abstract{Exotic quantum spin liquid (QSL) is formed with such
hypothetic particles as fermionic spinons carrying spin $1/2$  and
no charge. Here we calculate its thermodynamic and relaxation
properties. Our calculations unveil the fundamental properties of
QSL, forming strongly correlated Fermi system located at a fermion
condensation quantum phase transition. These are in a good
agreement with experimental data and allow us to detect the
behavior of QSL as that observed in heavy fermion metals. We
predict that the thermal resistivity of QSL under the application
of magnetic fields at fixed temperature demonstrates a very
specific behavior. The key features of our findings are the
presence of spin-charge separation and QSL formed with itinerant
heavy spinons in herbertsmithite.}

\begin{document}
\maketitle

A search for the materials formed with fermionic spinons carrying
spin $1/2$  and no charge is a challenge for a condensed matter
physics \cite{bal}. A quantum spin liquid (QSL) can be viewed as an
exotic quantum state of matter composed of hypothetic particles
like chargeless fermionic spinons with spin $1/2$ and having a
possibility to fluctuate as in ordinary fluid. The advances in
theoretical studies of QSLs date back to Anderson's seminal paper
\cite{and} which ignited hot discussions in the scientific
community. Theoretical consensus is that the ground state of QSLs
is not magnetically ordered, including both gapped and gapless spin
liquids \cite{bal}. However, the experimental investigation of QSLs
is hindered by a lack of real solids where it can occur. In other
words, no specific model substances with QSL has been found yet,
although there are few possible candidates. This means that the
recognition of a perfect material with clear QSL realization makes
a major challenge in modern condensed matter physics.

The experimental studies of herbertsmithite \chem{ZnCu_3(OH)_6Cl_2}
single crystal have found no evidences of long range magnetic order
or spin freezing indicating that \chem{ZnCu_3(OH)_6Cl_2} is the
promising system to investigate QSL
\cite{herb0,herb4,herb1,herb2,herb3,herb,sl,sl1,sl2,sl3}. Presently
herbertsmithite \chem{ZnCu_3(OH)_6Cl_2} has been exposed as a
$S=1/2$ Heisenberg antiferromagnet \cite{herb0} on a perfect kagome
lattice and new experimental investigations have revealed its
unusual behavior \cite{herb1,herb2,herb3,herb}, see Ref.
\cite{herb4} for a recent review. In \chem{ZnCu_3(OH)_6Cl_2}, the
$\rm Cu^{2+}$ ions with $S=1/2$ form the triangular kagome lattice,
and are separated by nonmagnetic intermediate layers of $\rm Zn$
and $\rm Cl$ atoms. The planes of the $\rm Cu^{2+}$ ions can be
considered as two-dimensional (2D) layers with negligible magnetic
interactions along the third dimension.

A frustration of simple kagome lattice leads to a dispersionless
topologically protected branch of the spectrum with zero excitation
energy known as the flat band \cite{green,vol}. In this  case the
fermion condensation quantum phase transition (FCQPT)
\cite{pr,prbr} can be considered as quantum critical point (QCP) of
the \chem{ZnCu_3(OH)_6Cl_2} QSL composed of chargeless fermions
with $S=1/2$ occupying the corresponding Fermi sphere with the
Fermi momentum $p_F$. As we are dealing with the 3D compound
\chem{ZnCu_3(OH)_6Cl_2} rather than with an ideal 2D kagome
lattice, we have to bear in mind that the real magnetic
interactions in the substance can shift the QSL from the exact
FCQPT point, positioning it in front of or behind the QCP.
Therefore, the actual location has to be established by analyzing
the experimental data only. It is believed that the $S=1/2$ model
on the kagome lattice can be viewed as a gapless spin liquid
\cite{herb1,herb2,herb3,herb,herb4,sl,sl1,sl2,sl3}, while recent
accurate calculations point to a fully gapped one, see
\cite{sci_slg} and Refs. therein. Thus, it is of crucial importance
to test what kind of QSL is formed in herbertsmithite and
determines its low temperature thermodynamic and relaxation
properties.

\begin{figure} [! ht]
\begin{center}
%\vspace*{-0.7cm}
\includegraphics [width=0.48\textwidth]{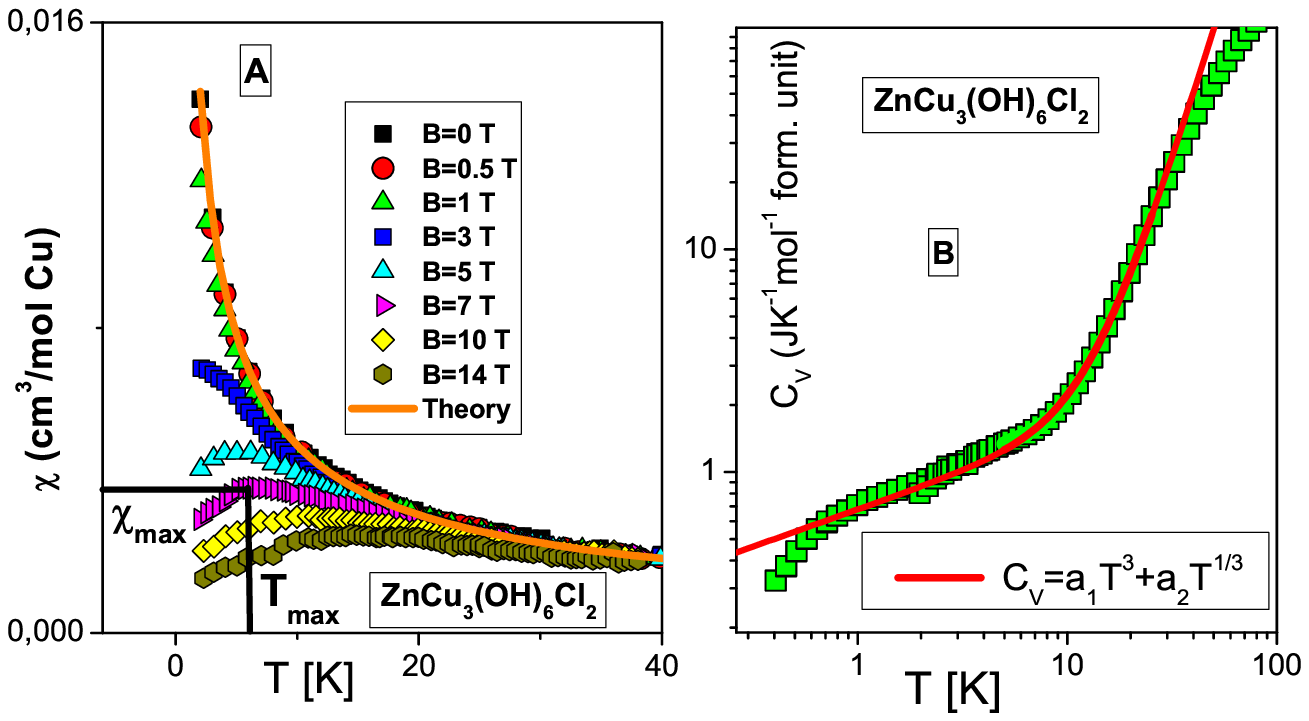}
\includegraphics [width=0.37\textwidth]{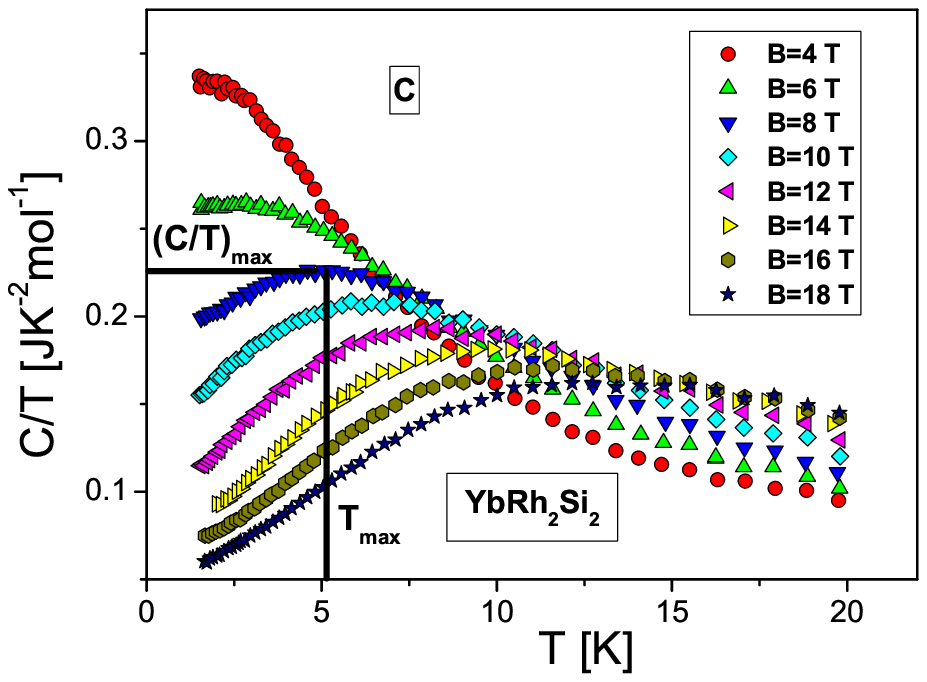}
%\vspace*{-0.5cm}
\end{center}
%\vspace*{-0.8cm}
\caption{Panel A: $T$-dependence of the magnetic susceptibility
$\chi$ at different magnetic fields $B$ \cite{herb3} shown in the
legend. The values of $\chi_{\rm max}$ and $T_{\rm max}$ at $B=7$ T
are also shown. Our calculations at $B=0$ are depicted by the solid
curve $\chi(T)\propto T^{-\alpha}$ with $\alpha=2/3$. Panel B: The
heat capacity measured on \chem{ZnCu_3(OH)_6Cl_2} at zero magnetic
field \cite{herb2} is shown by squares. Solid curve corresponds to
our theoretical approximation based on the function
$C=a_1T^3+a_2T^{1/3}$ with fitting parameters $a_1$ and $a_2$, see
eq. \eqref{CT}. Panel C reports the $T$-dependence of the
electronic specific heat $C/T$ of \chem{ YbRh_2Si_2} at different
magnetic fields \cite{steg1} as shown in the legend. The values of
$(C/T)_{\rm max}$ and $T_{\rm max}$ at $B=8$ T are also
shown.}\label{fig1}
\end{figure}
To identify unambiguously the compound having an exotic QSL located
at FCQPT, we must analyze a multitude of experimental features
rather than a single one. At very low temperatures and with
magnetic field $B$ applied, we expect that the possible QSL
substances reveal a Landau Fermi-liquid (LFL) behavior of their
physical characteristics, like the magnetic specific heat and
susceptibility, spin-lattice relaxation rate and the heat transport
coefficients. At the same time, at elevated temperatures a
non-Fermi liquid (NFL) regime emerges separated by a transition (or
crossover) region. In other words, the candidate substance should
exhibit the LFL, NFL and the transition regimes similar to the case
of heavy fermion (HF) metals and 2D $\rm ^3 He$ \cite{pr,
prl,lohn,prbr}. This means that QSL plays the role of HF liquid
placed into the insulating compound. Spin is carried by an electron
which becomes localized, while the spin is itinerant. Such a
behavior resembles the spin-charge separation observed recently in
1D system of interacting non-localized electrons \cite{scs}. In our
case the charges is completely localized while the spins are
itinerant dwelling on the topologically protected branch. As a
result, we face a breathtaking picture: New particles govern the
systems properties at low temperatures and these are different from
the particles presented in the Hamiltonian describing the system.

\begin{figure}[! ht]
\begin{center}
%\vspace*{-0.6cm}
\includegraphics [width=0.44\textwidth]{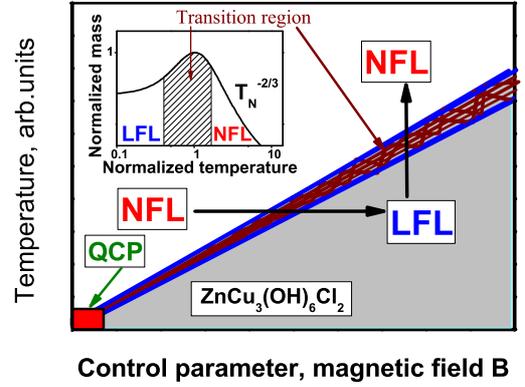}
%\vspace*{-1.0cm}
\end{center}
\caption{Schematic phase diagram of \chem{ZnCu_3(OH)_6Cl_2}. The
square at the origin shown by the arrow represents QCP near which
QSP is located. Vertical and horizontal arrows show LFL-NFL and
NFL-LFL transitions at fixed $B$ and $T$ respectively. At fixed $T$
the increase of $B$ drives the system along the horizontal arrow
from NFL regime to LFL one. On the contrary, at fixed $B$ and
raising temperatures the system transits along the vertical arrow
from LFL regime to NFL one. The inset demonstrates the behavior of
the normalized effective mass $M^*_N$ versus normalized temperature
$T_N$ as given by eq. \eqref{UN2}. It is seen that temperatures
$T_N\sim 1$ signify a transition region between the LFL regime with
almost constant effective mass and NFL one, given by $T^{-2/3}$
dependence. It is seen from eq. \eqref{UN2} that the width of the
transition region $T_w\propto T\propto B$. The transition region,
where $M^*_N$ reaches its maximum at $T/T_{\rm max}=1$, is shown by
the arrows and hatched area both in the main panel and in the
inset.}\label{fig0}
\end{figure}

We begin with the analysis of the magnetic susceptibility $\chi(T)$
of \chem{ZnCu_3(OH)_6Cl_2} shown in the panel A of fig. \ref{fig1}.
It displays an unusual behavior \cite{herb3}: at $B\geq 3$ T,
$\chi(T)$ has a maximum $\chi_{\rm max}(T)$ at some temperature
$T_{\rm max}(B)$. The maximum $\chi_{\rm max}(T)$ decreases as
magnetic field $B$ grows, while $T_{\rm max}(B)$ shifts to higher
$T$ reaching $15$ K at $B=14$ T.  It is seen from the panel A of
fig. \ref{fig1} that $\chi(T)\propto T^{-\alpha}$ with
$\alpha=2/3$. The calculated exponent is in good agreement with the
experimental value $\alpha=2/3\simeq 0.66$ \cite{herb3}.  The
observed behavior of $\chi$ strongly resembles that in HF metals
and associated with their proximity to FCQPT \cite{pr,prbr}. The
specific heat $C$, arising from the $\rm Cu$ spin system, at the
lowest explored temperatures, $106<T<400$ mK, follows a linear
temperature dependence, $C\propto T$. As it is seen from fig.
\ref{fig1}, panel B, for temperatures of a few Kelvin and higher,
the specific heat becomes $C(T)\propto T^3$ and is dominated by the
lattice contribution. At low temperatures $T\leq 1$ K, the strong
magnetic field dependence of the specific heat $C$ suggests that it
is predominately formed by the specific heat $C_{mag}$ of QSL,
$C_{mag}=C-a_1T^3$, since the lattice contribution is independent
of $B$ \cite{herb1,herb2,herb3}. The above behavior of $\chi$ (fig.
\ref{fig1}, panel A) is a visible parallel to that of the HF metal
\chem{ YbRh_2Si_2} observed in measurements of $C/T$ and displayed
in the panel C. This coincidence becomes evident if we recollect
that for HF liquid $\chi\propto C/T$ \cite{pr}. It is seen from the
panel C that the electronic specific heat of \chem{ YbRh_2Si_2}
\cite{steg1} is also strongly dependent on the applied magnetic
field and we will see below that both the specific heat $C_{mag}$
and that of \chem{ YbRh_2Si_2} exhibit the same behavior.

\begin{figure}[! ht]
\begin{center}
%\vspace*{-0.6cm}
\includegraphics [width=0.47\textwidth]{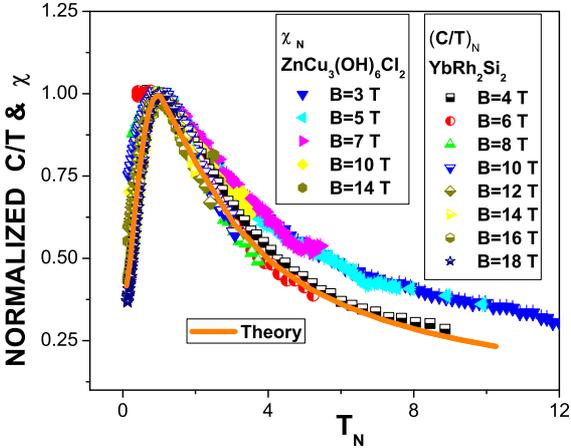}
%\vspace*{-1.0cm}
\end{center}
\caption{The experimental data on measurements of $\chi_N$
\cite{herb3}, $(C/T)_N=M^*_N$ \cite{steg1} and our calculations of
$M^*_N$ at fixed high magnetic field (high means that it polarizes
completely the quasiparticles spins) are shown by points of
different shape and solid curve respectively. It is clearly seen
that the data collected on both \chem{ZnCu_3(OH)_6Cl_2} and \chem{
YbRh_2Si_2} merge into the same curve, obeying the scaling
behavior. In accordance with the phase diagram displayed in the
panel A, at growing temperatures ($y\simeq 1$) the LFL regime first
converts into the transition one and then disrupts into the NFL
regime. This demonstrates that the spin liquid of herbertsmithite
is close to QCP and behaves like HF liquid of \chem{ YbRh_2Si_2} in
strong magnetic fields \cite{epl}.}\label{fig01}
\end{figure}
To study theoretically the low temperature thermodynamic,
relaxation and scaling properties of herbertsmithite, we use the
model of homogeneous HF liquid \cite{pr}. This model permits to
avoid complications associated with the crystalline anisotropy of
solids. In our case, similar to the above HF liquid, QSL is
composed of chargeless fermions (spinons) with $S=1/2$ occupying
the corresponding two Fermi spheres with the Fermi momentum $p_F$.
The ground state energy $E(n)$ is given by the Landau functional
depending on the quasiparticle distribution function $n_\sigma({\bf
p})$, where $p$ is the momentum and $\sigma$ is the spin index. The
effective mass $M^*$ is governed by the Landau equation
\cite{land,pr}
\begin{eqnarray}
\nonumber
&&\frac{1}{M^*(T,B)}=\frac{1}{M^*}+\frac{1}{p_F^2}\sum_{\sigma_1}
\int\frac{{\bf p}_F{\bf p_1}}{p_F}\\
&\times&F_{\sigma,\sigma_1}({\bf p_F},{\bf
p}_1)\frac{\partial\delta n_{\sigma_1}({\bf p}_1,T,B)}
{\partial{p}_1}\frac{d{\bf p}_1}{(2\pi) ^3}. \label{HC3}
\end{eqnarray}
Here we rewrite the quasiparticle distribution function as
$n_{\sigma}({\bf p},T,B) \equiv n_{\sigma}({\bf p},T=0,B=0)+\delta
n_{\sigma}({\bf p},T,B)$. The Landau amplitude $F$ is completely
defined by the fact that the system has to be at QCP of FCQPT
\cite{pr,ckz,epl,khodb}, see \cite{epl} for details of solving eq.
\eqref{HC3}. The sole role of Landau amplitude is to bring the
system to FCQPT point, where Fermi surface alters its topology so
that the effective mass acquires temperature and field dependencies.
At this point, the term $1/M^*$ vanishes, eq. \eqref{HC3} becomes
homogeneous and can be solved analytically \cite{pr,ckz}. At $B=0$,
the effective mass, being strongly $T$ - dependent, demonstrates the
NFL behavior given by eq. \eqref{HC3}
\begin{equation}
M^*(T)\simeq a_TT^{-2/3}.\label{MTT}
\end{equation}
At finite $T$, under the application of magnetic field $B$ the two
Fermi spheres due to the Zeeman splitting are displaced by opposite
amounts, the final chemical potential $\mu$ remaining the same
within corrections of order $B^2$. As a result, field $B$ drives
the system to LFL region, and again it follows from  eq.
\eqref{HC3} that
\begin{equation}
M^*(B)\simeq a_BB^{-2/3}. \label{MBB}
\end{equation}
It is seen from eqs. \eqref{MTT} and  \eqref{MBB} that effective
mass diverges at FCQPT. At finite $B$ and $T$ near FCQPT, the
solutions of eq. \eqref{HC3} $M^*(B,T)$ can be well approximated by
a simple universal interpolating function. This interpolation occurs
between the LFL regime, given by eq. \eqref{MBB} and NFL regime
given by eq. \eqref{MTT} \cite{pr,ckz}. In the case of strongly
correlated Fermi systems like HF metals and 2D $\rm ^3He$ the
thermodynamic and relaxation properties are defined by the effective
mass $M^*$, namely $\chi\propto(C/T)\propto\sqrt{1/T_1T}\propto
M^*$, where $1/T_1T\propto\chi^2$ is the spin-lattice relaxation
rate \cite{pr}. To study the universal scaling behavior of strongly
correlated Fermi system, it is convenient to introduce the
normalized effective mass $M^*_N$ and the normalized temperature
$T_N$ dividing the effective mass $M^*$ and temperature $T$ by their
maximal values, $M^*_M$ and $T_M$ respectively. The behavior of
$M^*_N=M^*/M^*_M$ as a function of $y=T/T_M$ shown in the inset to
fig. \ref{fig0} is independent of the specific features of
corresponding strongly correlated Fermi system, while both $M^*_M$
and $T_M$ are determined by these features \cite{pr}. As a result,
we obtain \cite{pr}
\begin{equation}
\chi_N=(C_{mag}/T)_N=\left(\sqrt{1/T_1T}\right)_N=M^*_N,
\label{SCB}
\end{equation}
where $\chi_N$, $(C/T)_N$ and $(\sqrt{1/T_1T})_N$ are the
normalized values of $\chi$, $C/T$ and $\sqrt{1/T_1T}$,
respectively. As a typical example, the corresponding maximum
values $\chi_{max}$, $(C/T)_{max}$ and $T_{max}$ used to normalize
the susceptibility $\chi$ and $C/T$ are shown in fig. \ref{fig1},
the panels A and C. We note that our calculations of $M^*_N$ based
on eq. \eqref{HC3} do not contain any fitting parameters. The
normalized effective mass $M^*_N=M^*/M^*_M$ as a function of the
normalized temperature $y=T_N=T/T_M$ reads
\begin{equation}M^*_N(y)\approx c_0\frac{1+c_1y^2}{1+c_2y^{8/3}}.
\label{UN2}
\end{equation}
Here $c_0=(1+c_2)/(1+c_1)$, $c_1$ and $c_2$ are fitting parameters,
approximating the Landau amplitude. Since magnetic field $B$ enters
eq. \eqref{HC3} only in combination $B\mu_B/T$, we have $T_{\rm
max}\propto B\mu_B$ \cite{ckz,pr}, where $\mu_B$ is the Bohr
magneton. Thus, for finite magnetic fields variable $y$ becomes
\begin{equation}\label{YTB}
y=T/T_{N}\propto T/\mu_BB.
\end{equation}
Subsequently we use eq. \eqref{UN2} to clarify our calculations
based on eq. \eqref{HC3}. The variables $B$ and $T$ enter eq.
\eqref{HC3} symmetrically, therefore eq. \eqref{UN2} determines
$M^*$ as a function of $B$ at fixed $T$ \cite{pr}. It follows
directly from eqs. \eqref{MBB}, \eqref{UN2} and \eqref{HC3} that
\begin{equation}\label{SCCHI}
\chi(T/\mu_BB)\,T^{2/3}\simeq \chi(T,B)\,T^{0.66}\propto
y^{2/3}M^*_N(y).
\end{equation}
Since the magnetization $M(T,B)=\int \chi(T,b)db$, we obtain that
\begin{equation}\label{SCM}
M(T,B)\,T^{-1/3}\simeq M(T,B)\,T^{-0.34}
\end{equation}
depends on the only variable $y$. Equations \eqref{SCCHI} and
\eqref{SCM} confirm the scaling behavior of both $\chi T^{0.66}$
and $M T^{-0.34}$ experimentally established in Ref. \cite{herb3}

\begin{figure}[! ht]
\begin{center}
%\vspace*{-0.6cm}
\includegraphics [width=0.47\textwidth]{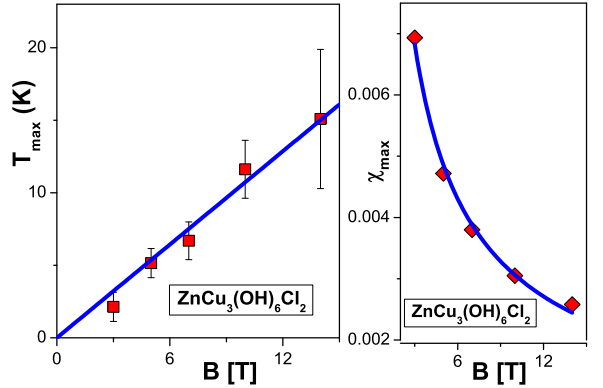}
%\vspace*{-1.0cm}
\end{center}
\caption{Left panel: The temperatures $T_{\rm max}(B)$ at which the
maxima of $\chi$ (see fig. \ref{fig1}, the panel A) are located.
The solid line represents the function $T_{\rm max}\propto aB$, $a$
is a fitting parameter, see eq. \eqref{YTB}. Right panel: the
maxima $\chi_{\rm max}$ of the functions $\chi(T)$ versus magnetic
field $B$ (see fig. \ref{fig1}, the panel A). The solid curve is
approximated by $\chi_{\rm max}(B)=dB^{-2/3}$, see eq. \eqref{MBB},
$d$ is a fitting parameter.}\label{fig02}
\end{figure}
Now we construct the schematic phase diagram of
\chem{ZnCu_3(OH)_6Cl_2}. This is reported in fig. \ref{fig0}. At
$T=0$ and $B=0$ the system is located near FCQPT without tuning. In
fact, it located in front of FCQPT since $C(T)$ demonstrates the
LFL behavior characterized by the linear $T$-dependence at low
temperatures \cite{herb1}. Both magnetic field $B$ and temperature
$T$ play role of the control parameters, driving the system from
NFL to LFL regions as shown by the vertical and horizontal arrows.
It follows from fig. \ref{fig01} that in accordance with eq.
\eqref{SCB} the behavior of $\chi_N$ coincides with that of
$(C/T)_N$ in \chem{ YbRh_2Si_2}. Figure \ref{fig02}, the right and
left panels correspondingly, demonstrates that eqs. \eqref{YTB} and
\eqref{MBB} are in a good agreement with the experimental facts.

\begin{figure} [! ht]
\begin{center}
%\vspace*{-0.7cm}
\includegraphics [width=0.47\textwidth]{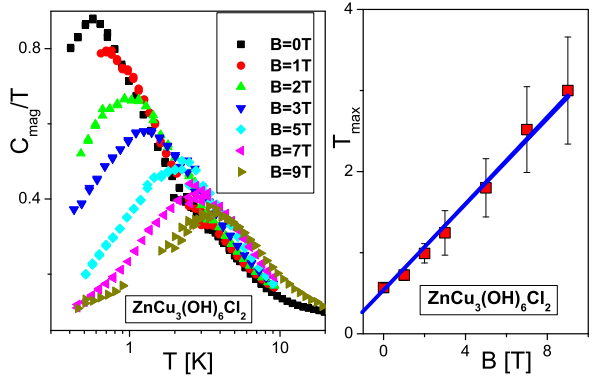}
%\vspace*{-0.5cm}
\end{center}
%\vspace*{-0.8cm}
\caption{Left panel: The specific heat $C_{mag}/T$ of QSL given by
eq. \eqref{CT} is extracted from measurements of $C(B)$ on
\chem{ZnCu_3(OH)_6Cl_2} at different magnetic fields shown in the
legend \cite{herb2}. Right panel: The temperatures $T_{\rm max}(B)$
at which the maxima of $C_{mag}/T$ (see the left panel) are located.
The solid line represents the function $T_{\rm max}\propto B$, see
eq. \eqref{YTB}.}\label{fig21}
\end{figure}

\begin{figure} [! ht]
\begin{center}
%\vspace*{-0.7cm}
\includegraphics [width=0.38\textwidth]{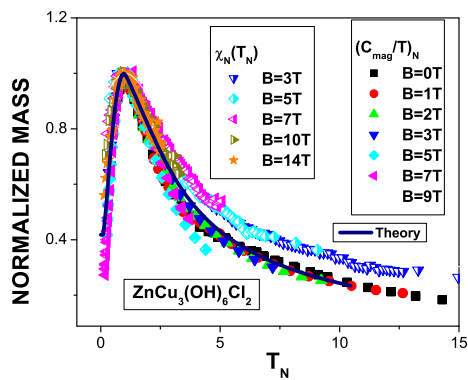}
%\vspace*{-0.5cm}
\end{center}
%\vspace*{-0.8cm}
\caption{The normalized susceptibility $\chi_N=\chi/\chi_{\rm
max}=M^*_N$ and the normalized specific heat $(C_{mag}/T)_N=M^*_N$
of QSL effective mass versus normalized temperature $T_N$ as a
function of the magnetic fields shown in the legends. $\chi_N$ is
extracted from the measurements of the magnetic susceptibility
$\chi$ in magnetic fields $B$ \cite{herb3} shown in the panel A of
fig. \ref{fig1}. The normalized specific heat is extracted from the
data displayed in fig. \ref{fig21}, the left panel. Our
calculations are depicted by the solid curve tracing the scaling
behavior of $M^*_N$.}\label{fig22}
\end{figure}

\begin{figure} [! ht]
\begin{center}
%\vspace*{-0.7cm}
\includegraphics [width=0.38\textwidth]{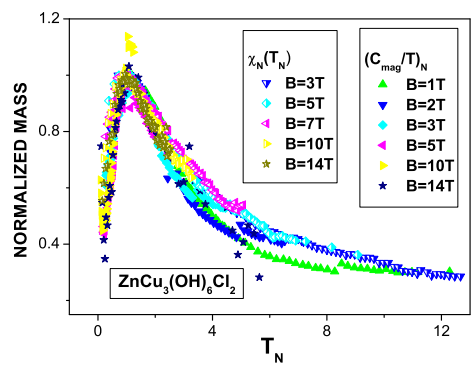}
%\vspace*{-0.5cm}
\end{center}
%\vspace*{-0.8cm}
\caption{The normalized susceptibility $\chi_N$ and the normalized
specific heat $(C_{mag}/T)_N$ of QSL effective mass versus
normalized temperature $T_N$ as a function of the magnetic fields
shown in the legends. $\chi_N$ and $(C_{mag}/T)_N$ are extracted
from the data of \cite{herb3} and \cite{herb1},
respectively.}\label{fig23}
\end{figure}

According to eq. \eqref{SCB}, in the case of QSL the behavior of the
specific heat $(C_{mag}/T)_N$ must coincide with that of $\chi_N$.
To separate $C_{mag}$ contribution, we approximate the general
specific heat $C(T)$ at $T>2$ K by the function
\begin{equation}\label{CT}
C(T)=a_1T^3+a_2T^{1/3},
\end{equation}
where the first term proportional to $a_1$ is due to the lattice
(phonon) contribution and the second one is determined by the QSL
when QSL exhibits the NFL behavior as it follows from eq.
\eqref{MTT}. It is seen from fig. \ref{fig1}, the panel B, that the
approximation is valid in the wide temperature range. We note that
the value of $a_1$ is almost independent of $a_2$, the presence of
which allows us to achieve a better approximation for $C$. The
obtained heat capacity $C_{mag}/T=(C-a_1T^3)/T$ is displayed in the
left panel of fig. \ref{fig21}, while the right panel B
demonstrates the maximum temperature as a function of the magnetic
field $B$. It is seen that $C_{mag}/T\propto M^*$ behaves like
$\chi\propto M^*$ shown in fig. \ref{fig1}, the panel A. In figs.
\ref{fig22} and \ref{fig23}, the normalized $(C_{mag}/T)_N$ and
$\chi_N$ are depicted. It is seen from both these figures that the
results obtained on different samples and measurements
\cite{herb1,herb2} exhibit similar properties. As it is seen from
figs. \ref{fig01}, \ref{fig22} and \ref{fig23} that in accordance
with eq. \eqref{SCB}, $(C_{mag}/T)_N\simeq\chi_N$ displays the same
scaling behavior as $(C/T)_N$ measured on the HF metal \chem{
YbRh_2Si_2}. Therefore, the scaling behavior of the thermodynamic
functions of herbertsmithite is the intrinsic feature of the
compound and has nothing to do with magnetic impurities. This
observation rules out a possible supposition that extra $\rm Cu$
spins outside the kagome planes considered as paramagnetic weakly
interacting impurities could be responsible for the divergent
behavior of the susceptibility at low temperatures seen from fig.
\ref{fig1}, the panel A. In that case the supposition is to lead to
explanations of the observed scaling behavior of $\chi$ and $C/T$
in magnetic fields shown in figs. \ref{fig01}, \ref{fig22} and
\ref{fig23}. Obviously, it is impossible.

\begin{figure} [! ht]
\begin{center}
%\vspace*{-0.7cm}
\includegraphics [width=0.36\textwidth]{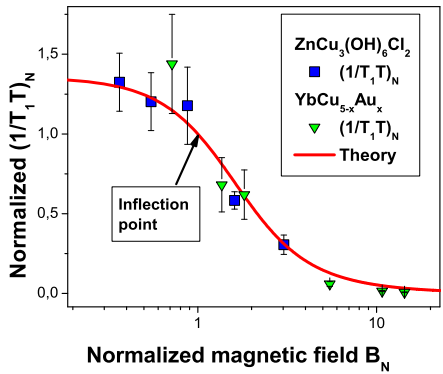}
%\vspace*{-0.5cm}
\end{center}
%\vspace*{-0.8cm}
\caption{The relaxation properties of herbertsmithite versus those
of HF metals. The normalized spin-lattice relaxation rate
$(1/T_1T)_N$ at fixed temperature as a function of magnetic field:
Solid squares correspond to data on $(1/T_1T)_N$ extracted from
measurements on \chem{ZnCu_3(OH)_6Cl_2} \cite{imai}, while the
solid triangles correspond to those extracted from measurements on
\chem{ YbCu_{5-x}Au_{x}} with $x=0.4$  \cite{carr}. The inflection
point where the normalization is taken is shown by the arrow. Our
calculations based on eq. \eqref{HC3} are depicted by the solid
curve tracing the scaling behavior of $(M^*_N)^2$.}\label{fig3}
\end{figure}

Figure \ref{fig3} displays the normalized spin-lattice relaxation
rates $(1/T_1T)_N$ at fixed temperature versus normalized magnetic
field $B_N$. We notice from fig. \ref{fig3} that the magnetic field
$B$ progressively reduces $1/T_1T$, and the spin-lattice relaxation
rate as a function of $B$ possesses an inflection point at some
$B=B_{inf}$ shown by the arrow. To clarify the scaling behavior in
that case, we normalize both the function $1/T_1T$ and the magnetic
field. Namely, we normalize $(1/T_1T)$ by its value at the
inflection point, and magnetic field is normalized by $B_{inf}$,
$B_N=B/B_{inf}$. It follows from eq. \eqref{SCB} that
$(1/T_1T)_N=(M^*_N)^2$ and we expect that different strongly
correlated Fermi systems located near FCQPT exhibit the same
behavior of the normalized spin-lattice relaxation rate \cite{pr}.
It is seen from fig. \ref{fig3}, that both herbertsmithite
\chem{ZnCu_3(OH)_6Cl_2} \cite{imai} and HF metal
\chem{YbCu_{5-x}Au_{x}} \cite{carr} demonstrate similar behavior of
the normalized spin-lattice relaxation rate. As seen from fig.
\ref{fig3}, at $B\leq B_{inf}$ (or $B_N\leq1$) the normalized
relaxation rate $(1/T_1T)_N$ depends weakly on the magnetic field,
while at higher fields $(1/T_1T)_N$ diminishes in agreement with
eq. \eqref{MBB}, $(1/T_1T)_N=(M^*_N)^2\propto B^{-4/3}$. Thus, in
accordance with the phase diagram shown in fig. \ref{fig0}, we
conclude that the application of magnetic field $B$ leads to
crossover from the NFL to LFL behavior and to the significant
reduction in the relaxation rate.

As it was mentioned above, QSL plays a role of HF liquid framed
into the insulating compound. Thus, we expect that QSL in
herbertsmithite behaves like the electronic liquid in HF metals if
the charge of an electron were zero. In that case, the thermal
resistivity $w=LT/\kappa$, $w=w_0+A_wT^2$, of QSL behaves like the
electrical resistivity $\rho=\rho_0+A_{\rho}T^2$ of the electronic
liquid, since $A_w$ represents the contribution of spinon-spinon
scattering to thermal transport, being analogous to the
contribution $A_{\rho}$ to charge transport by electron-electron
scattering. Here, $L$ is the Lorenz number, $\kappa$ is the thermal
conductivity of QSL, $\rho_0$ and $w_0$ are residual resistivity of
the electronic liquid and QSL, respectively, and coefficient
$A_w\sim A_{\rho}\propto (M^*)^2$ \cite{pr,khod}. Thus, we predict
that the coefficient $A_w$ of the thermal resistivity of QSL under
the application of magnetic fields at fixed temperature behaves
like the spin-lattice relaxation rate shown in fig. \ref{fig3},
$A_w(B)\propto 1/T_1T(B)\propto (M^*(B))^2$, while in the LFL
region at fixed magnetic fields the thermal conductivity is a
linear function of temperature, $\kappa\propto T$.

To conclude, we have calculated the thermodynamic and relaxation
properties of herbertsmithite. Our calculations unveil the
fundamental properties of QSL, forming strongly correlated Fermi
system located at FCQPT. These are in a good agreement with
experimental data and allow us to detect the behavior of QSL as that
observed in heavy fermion metals. The key features of our findings
are the presence of spin-charge separation and QSL formed with
itinerant heavy spinons in herbertsmithite. Herbertsmithite
represents a fascinating experimental example when the new
particles-spinons, non-existing as free ones, totally replace the
initial particles entering the Hamiltonian to dominate the
properties at low temperatures. It is highly desirable to perform
the detailed experimental studies of the transport properties of
herbertsmithite. Also, conditions for spinons to form a superfluid
liquid remain to be explored.

We are grateful to  V. A. Khodel for valuable discussions and to A.
Harrison and M. A. de Vries for their help in dealing with the
experimental data \cite{herb2}. This work was supported by U.S.
DOE, Division of Chemical Sciences, Office of Basic Energy
Sciences, Office of Energy Research, AFOSR and the RFBR \#
12-02-00017.


\begin{thebibliography}{99}

\bibitem{bal} \Name{Balents L.}
\REVIEW{Nature} {464}{2010}{199}.

\bibitem{and}  \Name{Anderson P. W.}
\REVIEW{Mater. Res. Bull.} {8}{1973}{153}.

\bibitem{herb0} \Name{Shores M. P., Nytko E. A., Bartlett B. M. \and
Nocera D. G.} \REVIEW{J. Am. Chem. Soc.} {127}{2005}{13462}.

\bibitem{herb1} \Name{Helton J. S.  {\it et al.}}
\REVIEW{Phys. Rev. Lett.} {98}{2007}{107204}.

\bibitem{herb2} \Name{de Vries M. A., Kamenev K. V., Kockelmann W. A., Sanchez-Benitez J.
 \and  Harrison A.} \REVIEW{Phys. Rev. Lett.} {100}{2008}{157205}.

\bibitem{herb3} \Name{Helton J. S.  {\it et al.}} \REVIEW{Phys. Rev. Lett.} {104}{2010}{147201}.

\bibitem{herb} \Name{Han T. H. {\it et al.}} \REVIEW{Phys. Rev. B} {83}
{2011}{100402(R)}.

\bibitem{herb4} \Name{Bert F. \and Mendels P.}  \REVIEW{J. Phys. Soc. Jpn.}
{79}{2010}{011001}.

\bibitem{sl} \Name{Lee S. \and Lee P. U.} \REVIEW{Phys. Rev. Lett.} {95}{2005}{036403}.

\bibitem{sl1}  \Name{Motrunich O. I.} \REVIEW{Phys. Rev. B} {72}
{2005}{045105}.

\bibitem{sl2}  \Name{Ran Y., Hermele M., Lee P. A. \and Wen X.-G.}
\REVIEW{Phys. Rev. Lett.} {98}{2007}{117205}.

\bibitem{sl3} \Name{Ryu S., Motrunich O. I., Alicea J. \and Fisher
M. P. A.} \REVIEW{Phys. Rev. B} {75}{2007}{184406}.

\bibitem{green}  \Name{Green D., Santos L. \and Chamon C.}
\REVIEW{Phys. Rev. B} {82}{2010}{075104}.

\bibitem{vol}  \Name{Heikkil\"a T. T., Kopnin N. B. \and Volovik G. E.}
\REVIEW{JETP Lett.} {94}{2011}{233}.

\bibitem{pr}  \Name{Shaginyan V. R., Amusia M. Ya., Msezane A. Z. and
Popov K. G.} \REVIEW{Phys. Rep.} {492}{2010}{31}.

\bibitem{prbr}  \Name{Shaginyan V. R., Msezane A. Z. and
Popov K. G.} \REVIEW{Phys. Rev. B} {84}{2011}{060401(R)}.

\bibitem{sci_slg}  \Name{Yan S., Huse D. A. \and White S. R.} \REVIEW{Science}
{332}{2011}{1173}.

\bibitem{prl} \Name{Shaginyan V. R., Msezane A. Z., Popov K. G.
\and Stephanovich V. A.} \REVIEW{Phys. Rev. Lett.} {100}{2008}
{096406}.

\bibitem{lohn}  \Name{v. L\"ohneysen H., Rosch A., Vojta M. \and W\"olfle
P.}  \REVIEW{Rev. Mod. Phys.} {79}{2007}{1015}.

\bibitem{scs}  \Name{Jompol Y. {\it et al.}} \REVIEW{Science} {325}{2009}{597}.

\bibitem{steg1}  \Name{Gegenwart P. {\it et al}.}
\REVIEW{New J. Phys.} {8}{2006}{171}.

\bibitem{land}  \Name{Landau L. D.}
\REVIEW{Sov. Phys. JETP} {3}{1956}{920}.

\bibitem{ckz}  \Name{Clark J. W., Khodel V. A. \and Zverev M. V.}
\REVIEW{Phys. Rev. B} {71}{2005}{012401}.

\bibitem{khodb}  \Name{Khodel V. A., Clark J. W. \and Zverev M. V.}
\REVIEW{Phys. Rev. B} {78}{2008}{075120}.

\bibitem{epl}  \Name{Shaginyan V. R., Popov K. G., Stephanovich
V. A., Fomichev V. I. \and Kirichenko E. V.} \REVIEW{Europhys.
Lett.} {93}{2011}{17008}.

\bibitem{imai}  \Name{Imai T., Nytko E. A., Bartlett B. M., Shores M. P. \and Nocera
D. G.} \REVIEW{Phys. Rev. Lett.} {100}{2008}{077203}.

\bibitem{carr}  \Name{Carretta P., Pasero R., Giovannini M. \and Baines C.}
\REVIEW{Phys. Rev. B} {79}{2009}{020401(R)}.

\bibitem{khod}  \Name{Khodel V. A. and Schuck P.}  \REVIEW{Z. Phys. B} {104}{1997}{505}.

\end{thebibliography}
\end{document}